\documentclass[final,5p,twocolumn]{elsarticle}

\usepackage{amssymb}

\begin{document}

\begin{frontmatter}



\title{Event-by-event viscous hydrodynamics for Cu-Au collisions at $\sqrt{s_{NN}}=200$GeV}
\author{Piotr Bo\.zek}
\ead{piotr.bozek@ifj.edu.pl}
\address{The H. Niewodnicza\'nski Institute of Nuclear Physics,
PL-31342 Krak\'ow, Poland\\
Institute of Physics, Rzesz\'ow University, 
PL-35959 Rzesz\'ow, Poland}


\begin{abstract}
Event-by-event hydrodynamics is applied to  Cu-Au collisions at $\sqrt{s_{NN}}=200$GeV. 
Predictions for charged particle
 distributions in pseudorapidity, transverse momentum spectra, femtoscopy radii are given. The triangular 
and elliptic flow coefficients are calculated.
The directed flow at central rapidity in the reaction plane in asymmetric collisions is nonzero, 
fluctuations of the  initial profile lead to a further increase of the directed flow when  
measured in the event plane.
\end{abstract}


\end{frontmatter}








\bibliographystyle{model1a-num-names}

\biboptions{sort&compress}

The emission of particles with soft momenta in relativistic heavy-ion collisions can be described 
in a satisfactory way within the relativistic hydrodynamic model 
\cite{Kolb:2003dz,Florkowski:2010zz}. The harmonic flow coefficients are found to be sensitive to 
the viscosity of the expanding fluid \cite{Romatschke:2009im,Dusling:2007gi,Song:2011qa,Niemi:2011ix,Bozek:2009dw}.
Fluctuations determine the asymmetry of the initial distribution in the transverse plane 
\cite{Alver:2008zz,Alver:2010gr} and event-by-event hydrodynamics simulations are needed to 
investigate their effect on final spectra
 \cite{Andrade:2006yh,Gardim:2011qn,Bozek:2012fw,Schenke:2010rr,Qiu:2011hf,Chaudhuri:2011qm}. Typically collisions of symmetric nuclei are studied experimentally, the most important 
parameters varied in these analyses are the center of mass energy, the size of the system and the centrality of the 
collision. 

An additional insight into the mechanism of particle production in relativistic heavy-ion collisions can 
be gained from interactions of two different nuclei, e.g. Cu-Au collisions at $\sqrt{s_{NN}}=200$GeV.
The study of the multiplicity distribution and correlations in the longitudinal direction in asymmetric collisions gives additional constraints on 
 the mechanism of the energy deposition 
in the early stage of the reaction \cite{Gavin:2008ev,Gelis:2009tg,Dusling:2012ig,Bialas:2004su,Bialas:2011vj}, in particular 
models assuming approximate boost-invariance of the initial state must revised.
The  directed flow 
\cite{Bozek:2010bi,Andrade:2010zz,Csernai:1999nf,Snellings:1999bt} 
in asymmetric collisions is partly generated by a new mechanism, 
as discussed latter in this Letter. The measurement of differences between the directed flow of proton and antiprotons could give informations on the different 
rate of baryon stopping on the Cu and Au side of the fireball which could 
be compared to  model predictions
 \cite{Vance:1998vh,Weber:2002pq,MehtarTani:2008qg}.
The reaction plane correlations in the fireball 
\cite{Bozek:2010vz,Bhalerao:2011bp} are modified in asymmetric collisions, 
with stronger correlations between  odd and even order event planes.
 As the size and the energy density in the fireball are sufficient to 
allow for a substantial phase of collective expansion, we apply  the
 event-by-event viscous hydrodynamic model to obtain
 predictions for basic  observables for particles of soft momenta
produced in asymmetric Cu-Au interactions.

\begin{figure}[tb]
\begin{center}
\includegraphics[angle=0,width=0.47 \textwidth]{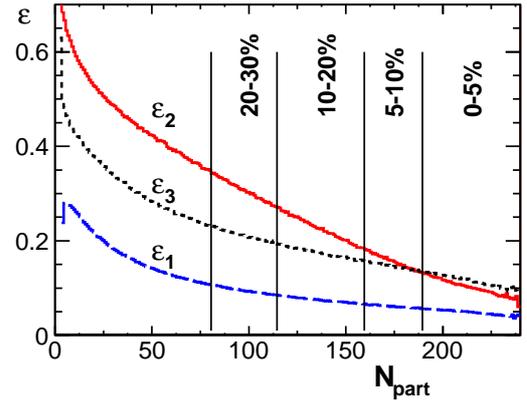} 
\end{center}
\caption{The eccentricity $\epsilon_2$ (solid line), triangularity $\epsilon_3$ 
(dotted line) and the dipole asymmetry $\epsilon_1$ (dashed line)
 of the initial fireball as function of the number 
of participant nucleons in Cu-Au collisions from  the Glauber Monte Carlo model 
\cite{Broniowski:2007nz}. The
vertical lines separate different  centrality classes. 
\label{fig:e2nw}} 
\end{figure}

The second order viscous hydrodynamic equations are solved in $3+1$-dimensions 
\cite{Schenke:2010rr,Bozek:2011ua}. 
We use two values for the  shear 
viscosity coefficient  $\eta/s=0.08$ or $0.16$ (unless stated otherwise the default value is $0.08$), the bulk viscosity is $\zeta/s=0.04$ in the 
hadronic phase.
At freeze-out particle emission is performed using a Monte Carlo procedure
\cite{Chojnacki:2011hb}, with a freeze-out temperature of $150$MeV. This freeze-out temperature gives a satisfactory description of  particle spectra in Au-Au collisions 
\cite{Bozek:2012fw}.
The hydrodynamic expansion is followed in each event with random initial density. The
initial density is generated from the Glauber Monte Carlo model \cite{Broniowski:2007nz}
with a Gaussian wounding profile for nucleon-nucleon collisions \cite{Rybczynski:2011wv}.
The entropy density in the transverse plane $x$-$y$ and space-time rapidity $\eta_\parallel$ is
given as a sum over the participant nucleons
\begin{eqnarray}
s(x,y,\eta_\parallel)&=& d \sum_i  f_\pm(\eta_\parallel) g_i(x,y) \left[
(1-\alpha)   
 + N^{coll}_i\alpha \right]. 
\label{eq:em}
\end{eqnarray}
weighted with  Gaussians $g_i(x,y)$ of width $0.4$fm centered at the positions of the 
participant nucleons $i$, a term  from  binary collisions  with $\alpha=0.125$ is included, 
 $N^{coll}_i$ is the number of collisions of the participant nucleon $i$.
The form of the longitudinal profile $f_{\pm}$ and the factor $d=2.5$~GeV are taken the
same as for Au-Au collisions at $200$GeV \cite{Bozek:2012fw}. 
In Fig. \ref{fig:e2nw} are shown the initial eccentricity $\epsilon_2$, triangularity $\epsilon_3$ and dipole asymmetry $\epsilon_1$ 
of the fireball as function of the number of participants,  averaged over the initial density with weights $r^2$, $r^3$ and $r^3$ respectively. 
In the following we present 
results for two centrality classes $0$-$5$\% and $20$-$30$\% defined by the number 
of participants $190 \le N_{part}$ and
$81 \le N_{part} \le 114$ respectively.  For each centrality class considered we generate $100$ hydrodynamic events, 
and for each hydrodynamically generated freeze-out 
configuration $1000$-$1500$
statistical emission events. The relative statistical error on $v_2$ and $v_3$ from 
sampling the initial eccentricity and triangularity distributions with $100$ events is 
$4-5$\%.

\begin{figure}[h]
\begin{center}
\includegraphics[angle=0,width=0.47 \textwidth]{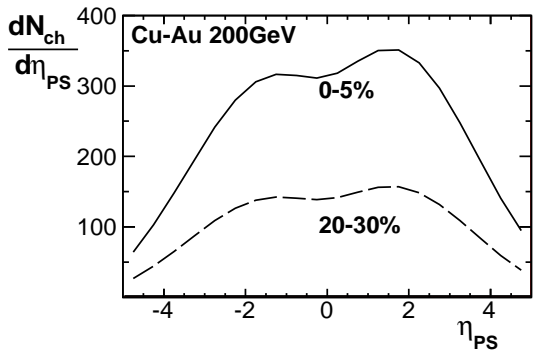} 
\end{center}
\caption{Distribution of charged particles in pseudorapidity for $0$-$5$\% (solid line) and $20$-$30$\% 
(dashed line) centrality classes. The Au momentum is directed
towards positive rapidity.
\label{fig:deta}} 
\end{figure}  
\begin{figure}[h]
\begin{center}
\includegraphics[angle=0,width=0.35 \textwidth]{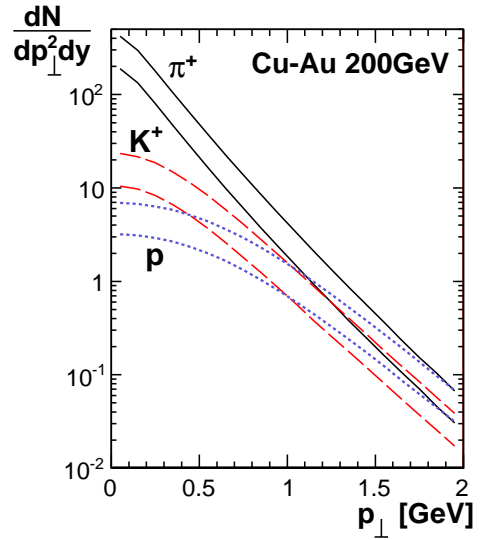} 
\end{center}
\caption{Transverse momentum spectra of $\pi^+$ (solid lines), $K^+$ (dashed lines) and
 protons (dotted lines) for $0$-$5$\% and 
$20$-$30$\% centrality classes (upper and lower curves respectively).
\label{fig:spectra}} 
\end{figure}

The distribution of charged particles in pseudorapidity $dN_{ch}/d\eta_{PS}$ is asymmetric 
(Fig. \ref{fig:deta}). It reflects 
the asymmetry of the initial entropy density (Eq. \ref{eq:em}) \cite{Bialas:2004su,Bozek:2011if}. 
The number of participant nucleons is 
larger in the Au nucleus, with the asymmetry increasing for central collisions. 
Using the initial condition extrapolated from Au-Au interactions at the same energy, the predicted
 multiplicities at central 
rapidity  are $dN_{ch}/d\eta_{PS}=320$ and
$140$ in $0$-$5$\% and $20$-$30$\% centrality classes respectively.

The transverse momentum spectra of $\pi^+$, $K^+$ and protons shown in Fig. \ref{fig:spectra} are 
softer than in Au-Au collisions. The predicted average transverse momenta of pions, 
kaons and protons in central collisions
are $405$, $587$ and $775$MeV respectively.
To obtain correct particle ratios nonzero 
chemical potentials are introduced at freeze-out, although the equation of 
state used to calculated the  dynamics of the system is at zero baryon density. 
 The equation of state is moderately 
changing with $\mu$ 
at small baryon densities \cite{Bluhm:2007nu}, and the equation of 
state at zero
baryon density  can be used for central rapidities in  collisions at $200$GeV. The  
the energy in the Cooper-Frye formula at freeze-out is conserved to 
the order $\mu^2/T^2$. 
The baryon chemical potential $\mu_B=22$MeV at the freeze-out temperature $150$MeV assures the 
same ratio $\bar{p}/{p}$ as measured for Au-Au interactions  \cite{Bearden:2003fw}.

\begin{figure}[bth]
\begin{center}
\includegraphics[angle=0,width=0.47 \textwidth]{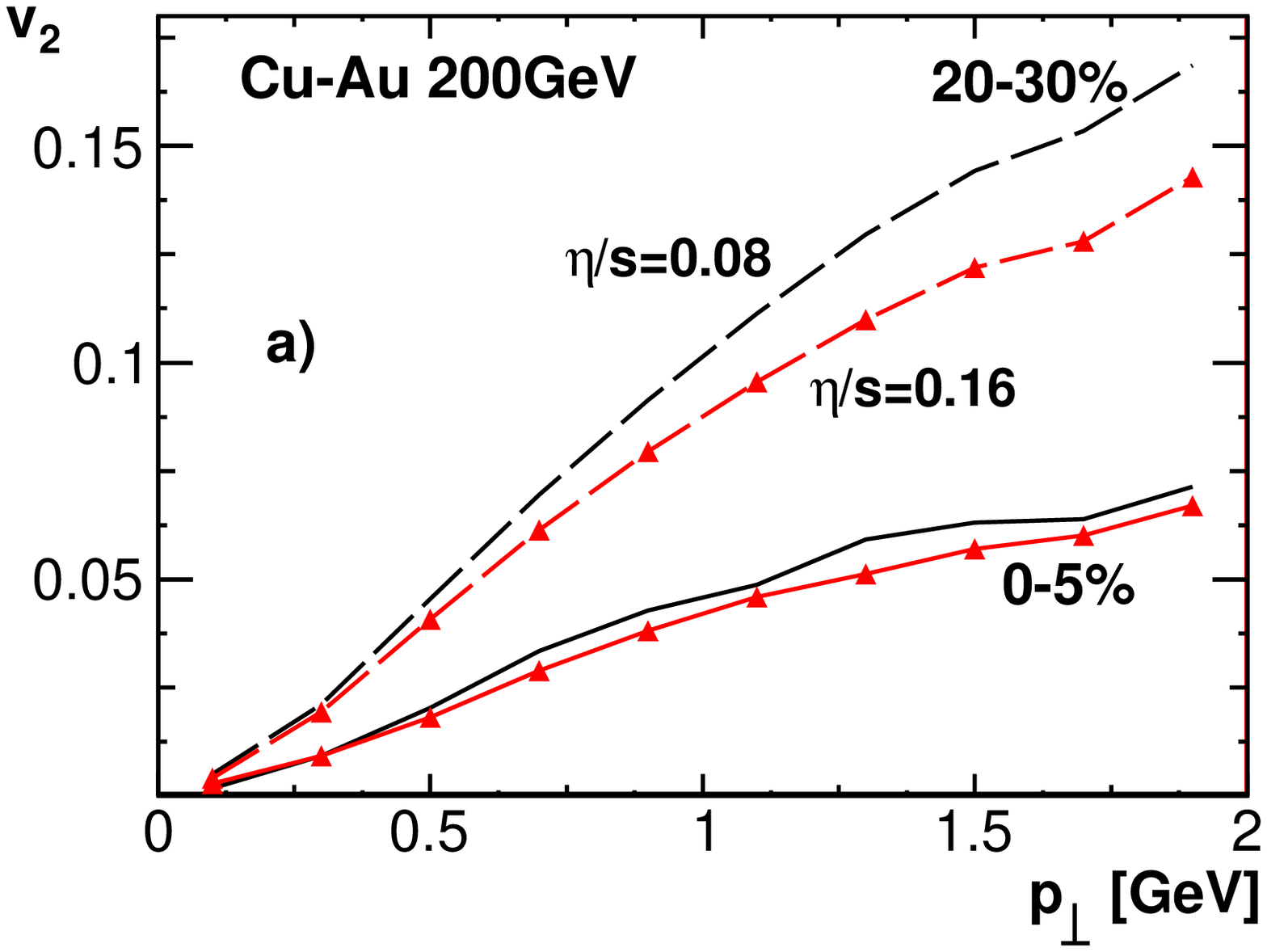} 
\includegraphics[angle=0,width=0.47 \textwidth]{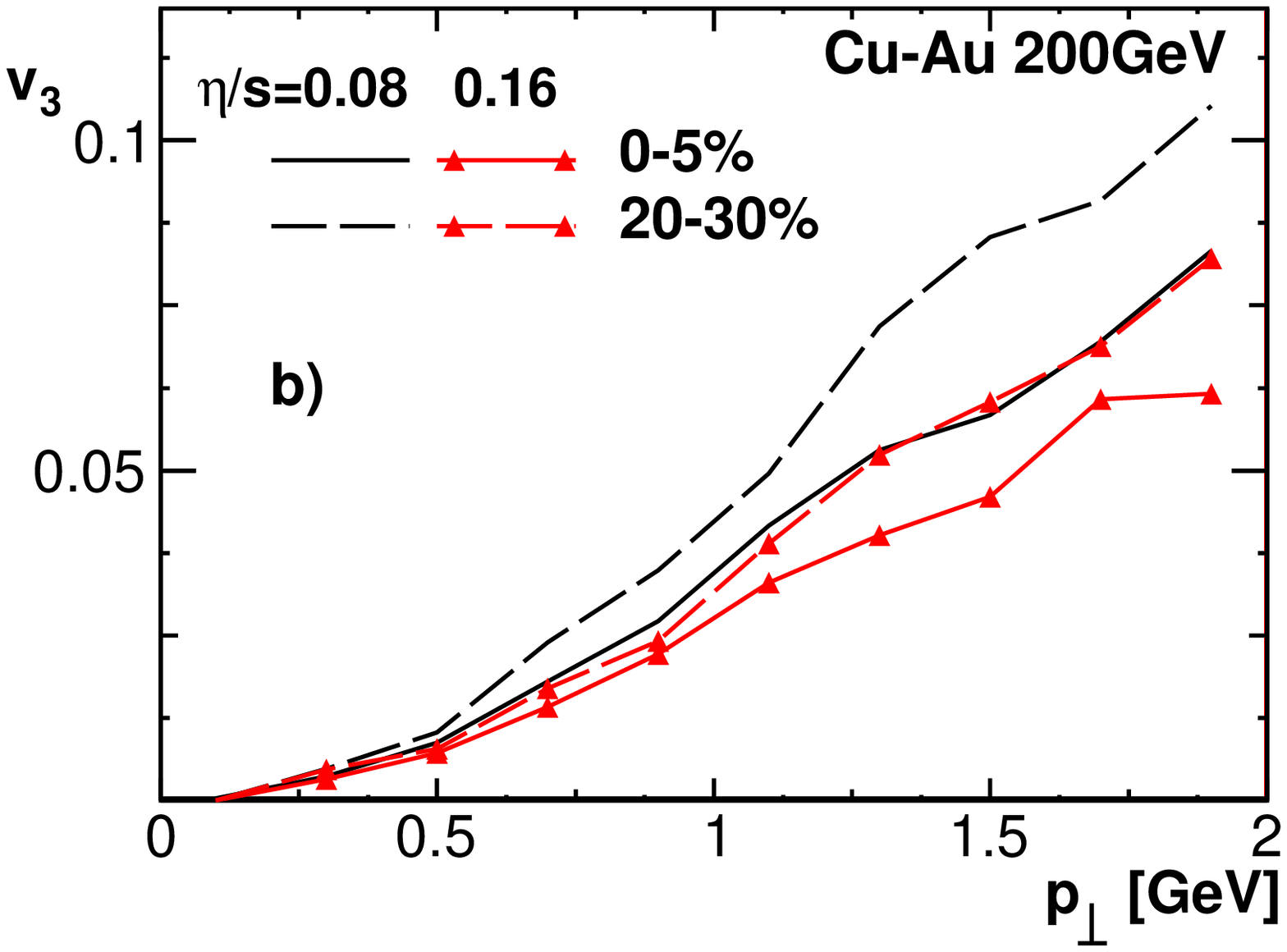} 
\end{center}
\caption{Elliptic (panel a) and triangular (panel b) flow coefficients of charged particles 
as function of transverse momentum   for centralities $0$-$5$\% (solid lines) and $20$-$30$\% (dashed lines).
The results of the calculation with $\eta/s=0.08$ are presented with lines and for $\eta/s=0.16$ using lines with triangles.
\label{fig:v23}} 
\end{figure}

The elliptic and triangular flow coefficients as function of $p_\perp$ are presented in Fig. \ref{fig:v23}. 
The differential flow coefficients are calculated using the two-particle cumulant method 
\cite{Borghini:2000sa}.  The elliptic flow coefficient for semi-central events 
is significantly larger than in central events. It reflects the  larger  value of
 the initial eccentricity in    the $20$-$30$\% centrality class (Fig. \ref{fig:e2nw}).
The triangular flow is similar in the two centrality classes studied.
With increasing value of  shear viscosity the elliptic and triangular asymmetry coefficients are reduced
\cite{Luzum:2008cw,Schenke:2010rr}. The influence of shear viscosity is larger in peripheral events
and for the triangular flow $v_3$ \cite{Song:2008si,Alver:2010dn}. The integrated value of the elliptic flow of charged particles 
in the $p_\perp$  range $150$-$2000$MeV is $0.022$ $(0.020)$ for central  and $0.048$ $(0.043)$ for peripheral events when $\eta/s=0.08$ $(0.16)$. 
The corresponding $v_3$ coefficients for charged particles 
are $0.013$ $(0.011)$ and $0.016$ 
$(0.012)$.

\begin{figure}[h]
\begin{center}
\includegraphics[angle=0,width=0.5 \textwidth]{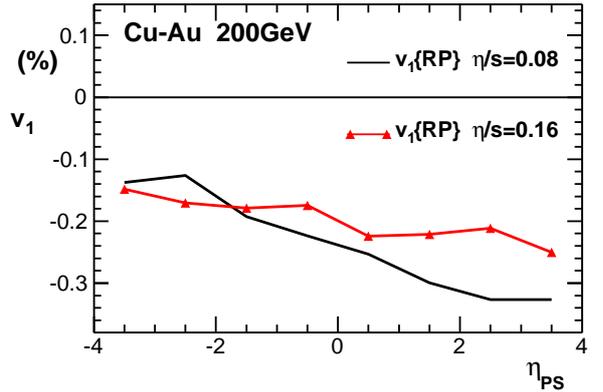} 
\end{center}
\caption{Directed flow of charged particles as function of pseudorapidity with respect to the reaction plane.
The results of the calculation with $\eta/s=0.08$ are presented using a solid line and for $\eta/s=0.16$ using a solid
line with triangles.
\label{fig:v1eta}} 
\end{figure}  

\begin{figure}[h]
\begin{center}
\includegraphics[angle=0,width=0.5 \textwidth]{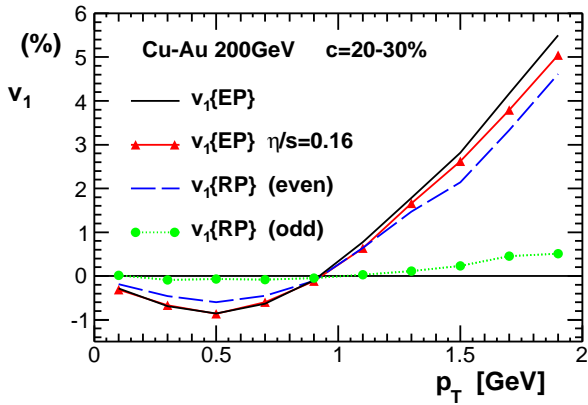} 
\end{center}
\caption{Directed flow of charged particles as function of transverse momentum
 with respect to the reaction plane (dashed line) and the event plane (solid lines) 
for $|\eta_{PS}|<1$, the dots represent the directed flow with respect to the reaction plane  odd in pseudorapidity for $|\eta_{PS}|<2$.
The results of the calculation of $v_1\{EP\}$
  with $\eta/s=0.16$ are presented  using a solid
line with triangles.
\label{fig:v1pt}} 
\end{figure}  

In collisions of symmetric nuclei at ultrarelativistic energies
 the directed flow exhibits a component odd in pseudorapidity
\cite{Back:2005pc,Abelev:2008jga,Bozek:2010bi,Andrade:2010zz}. At 
central rapidities nonzero directed flow occurs due to event-by-event 
fluctuations of the density profile \cite{Teaney:2010vd,Gardim:2011qn,ATLAS:2012at,Selyuzhenkov:2011zj}.
In the fireball created in the collision of two asymmetric nuclei such as 
Cu-Au collisions the profile is asymmetric in the 
reaction plane also on average. When 
defining the reaction plane always with the 
Au nucleus in the $x>0$ half plane  a nonzero component 
of the directed flow with respect to the reaction plane appears.

In Fig. \ref{fig:v1eta} is shown the $p_T$ integrated 
directed flow coefficient $v_1$ 
of charged particles 
with respect to the 
reaction plane as function of pseudorapidity for centrality $20$-$30$\%. 
At central rapidities the directed flow is negative,
more particles flow in the direction of the Cu half plane. 
Besides the nonzero component even in $\eta_{PS}$ a smaller  odd component of 
$v_1$ is visible in Fig. \ref{fig:v1eta}. In the hydrodynamic model the 
odd component of the directed flow is generated from the expansion of 
the fireball tilted away from the collision axis 
\cite{Bozek:2010bi,Andrade:2010zz}. The odd component of the directed flow is 
significantly reduced when  viscosity increases, this effect comes from the 
reduction of the longitudinal pressure in the early phase of the collisions 
for higher viscosity \cite{Bozek:2010aj}.

The even component of the 
directed flow as function of transverse momentum in the reaction plane 
\begin{equation}
v_1(p_\perp)\{RP\}= \langle \cos(\phi_i-\Psi_{RP})  \rangle
\end{equation} 
is estimated as the average for charged particles with $|\eta_{PS}|<1$. 
The direction of the reaction plane $\Psi_{RP}$ is chosen in the direction of the 
Au nucleus. Defining the reaction plane using all the nucleons in the 
Au and Cu nuclei or only the spectator nucleons in the two nuclei in each event
gives indistinguishable results. The directed flow $v_1(p_\perp)$
is negative for small momenta and changes sign for
$p_\perp\simeq 850$MeV (Fig. \ref{fig:v1pt}).  
The component of the directed flow odd in 
pseudorapidity
\begin{equation}
v_1(p_\perp)\{RP\}(odd)= \langle  {\textrm sgn}(\eta_{PS})
\cos(\phi_i-\Psi_{RP}) \rangle
\end{equation} 
is much smaller than the even one  
(we take charged particles with $|\eta_{PS}|<2$).  

Fluctuations of the fireball density in each event change the 
orientation and the magnitude of the directed flow in each event \cite{Teaney:2010vd}. 
We follow the procedure of Refs. \cite{Luzum:2010fb,Gardim:2011qn}, where the $Q$ vector 
of the event plane is defined with a weight reducing the contribution of 
momentum conservation to the directed flow
\begin{equation}
Q e^{i \Psi_1}= \langle w_i e^{i \phi_i} \rangle
\end{equation}
with $w_i = p_\perp -\langle p_\perp^2 \rangle /\langle p_\perp\rangle$.
The $Q$ weighted value of the directed flow coefficient is
\begin{equation}
v_1(p_\perp)\{EP\}=\frac{\langle Q \cos(\phi_i -\Psi_1)\rangle}{\sqrt{\langle Q^2 \rangle}}
\end{equation}
The directed flow coefficient $v_1(p_\perp)$ with respect to the event plane has 
the same form as the even component  defined in the reaction plane, but with a
slightly larger magnitude (solid line in Fig. \ref{fig:v1pt}). This means that fluctuations increase the directed 
flow at central rapidity. The calculated directed flow in the
 event plane 
does not depend 
strongly on the value of shear viscosity.

\begin{figure}[h]
\begin{center}
\includegraphics[angle=0,width=0.35 \textwidth]{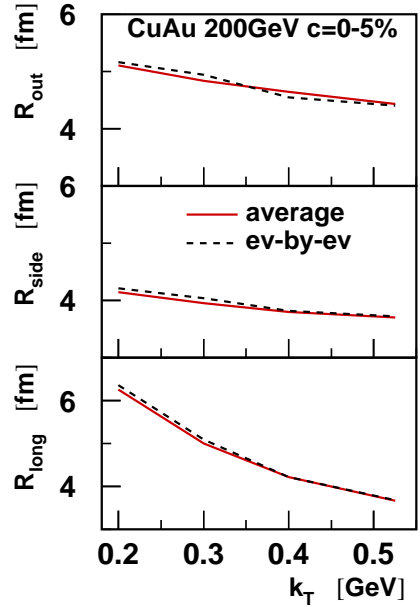} 
\end{center}
\caption{The femtoscopy radii $R_{out}$, $R_{side}$, and $R_{long}$ as function of the average transverse 
momentum of the pion pair. The dashed lines represent the results of the event-by-event hydrodynamics and the 
solid  lines the results obtained in a simulation using one average initial condition.
\label{fig:hbt}} 
\end{figure}

The correlation function for same charge 
pion  pairs is calculated using the momenta 
and positions of pions emitted in each event 
\cite{Kisiel:2006is,Chojnacki:2011hb}. From the correlation function in relative momentum of the pair the femtoscopy radii are extracted. In Fig. \ref{fig:hbt}
is shown the dependence of the three radii $R_{out}$, $R_{side}$, and $R_{long}$
on the average pion momentum. This dependence is similar as 
seen in Au-Au collisions, but with smaller values of the radii. In addition, we calculate
the femtoscopy radii using one hydrodynamic simulation starting from the 
average initial condition corresponding to  centrality $0$-$5$\% 
(solid lines in Fig \ref{fig:hbt}). The results are very similar
 as obtained in event by event simulations. It shows that flow fluctuations 
in the event by event evolution are too small to affect significantly the 
femtoscopy radii.

We present predictions of the event by event viscous hydrodynamic model for 
Cu-Au collisions at $\sqrt{s_{NN}}=200$GeV.
The charged particle multiplicity and femtoscopy radii 
are smaller than in  Au-Au interactions, reflecting the reduced size of the 
system. The transverse momentum spectra are steeper and the elliptic and 
triangular flow smaller than for the Au-Au system.

A novel aspect for collisions of asymmetric nuclei is seen
 in the directed flow. The asymmetry of the fireball density in the reaction 
plane leads to the formation of a directed flow of charged particles at 
central rapidities. This contribution appears when the orientation 
of the pair of Cu and Au nuclei 
in the reaction plane is  determined in each event and is nonzero 
also in the absence of density fluctuations.
Density fluctuations further increase the directed flow at central rapidities when measured in the 
reaction plane.

{\bf Acknowledgment: } 

Supported by Polish Ministry of Science and Higher Education, 
grant N~N202~263438.

\bibliography{../hydr}







\end{document}